\documentclass[runningheads]{llncs}

\usepackage[T1]{fontenc}
\usepackage{graphicx}
\graphicspath{{img/}} 
\usepackage[numbers,square,sort&compress]{natbib}
\usepackage[skip=0pt]{caption}
\usepackage{booktabs}
\usepackage{numprint}
\npthousandsep{,}
\npdecimalsign{.}
\usepackage[inline]{enumitem}
\usepackage{balance}
\usepackage{times}

\usepackage{xcolor}
\usepackage{verbatim}

 \usepackage{multirow}

\usepackage{acronym}
\usepackage{hyperref}

\usepackage{url}
\usepackage{xurl}

\usepackage{mathrsfs}

\usepackage{amsfonts}
\usepackage{amsmath}
\usepackage{amssymb}
\usepackage{bm,dsfont}
\usepackage{makecell}
\usepackage{hhline}
\usepackage{xcolor}
\usepackage{colortbl}
\usepackage[normalem]{ulem}
\usepackage{caption}
\usepackage{subcaption}
\usepackage{wrapfig}

\usepackage{pgfplotstable}
\usepackage{pgfplots}

\definecolor{lblue}{HTML}{A6CEE3}
\definecolor{lgreen}{HTML}{B2DF8A}
\definecolor{lred}{HTML}{FB9A99}
\definecolor{lorange}{HTML}{FDBF6F}
\definecolor{mblue}{HTML}{80B1D3}
\definecolor{mgreen}{HTML}{B3DE69}
\definecolor{mred}{HTML}{FB8072}
\definecolor{morange}{HTML}{FDB462}
\definecolor{blue}{HTML}{1F78B4}
\definecolor{green}{HTML}{33A02C}
\definecolor{red}{HTML}{E31A1C}
\definecolor{orange}{HTML}{FF7F00}
\definecolor{dblue}{HTML}{08519C}
\definecolor{dgreen}{HTML}{006D2C}
\definecolor{dorange}{HTML}{EC7014}

\newcommand{\header}[1]{\vspace{1mm}\noindent\textbf{#1.}}

\newcommand{\ourtask}{cross-modal retrieval}
\newcommand{\Ourtask}{Cross-modal retrieval}
\newcommand{\OURTASK}{Cross-modal Retrieval}
\newcommand{\ic}{image-caption pair}

\acrodef{SOTA}{state-of-the-art}
\acrodef{ViT}{Vision Transofrmer}
\acrodef{IR}{information retrieval}
\acrodef{CNN}{convolutional neural network}
\acrodef{CV}{computer vision}
\acrodef{CMR}{cross-modal retrieval}
\acrodef{RNG}{random number generator}

\acrodef{MLF}{Multi-level Feature approach}
\acrodef{CLIP}{Contrastive Language-Image Pretraining}

\acrodef{ABO}{Amazon Berkley Objects}
\acrodef{CUB-200}{Caltech-UCSD Birds 200}
\acrodef{MS COCO}{Microsoft COCO}
\newcommand{\coco}{\ac{MS COCO}}
\newcommand{\flickr}{Flickr30k}
\newcommand{\cub}{\ac{CUB-200}}
\newcommand{\abo}{\ac{ABO}}
\newcommand{\fashion}{Fashion200k}
\newcommand{\cocof}{\acf{MS COCO}~\cite{lin2014microsoft}}
\newcommand{\cubf}{\acf{CUB-200}~\cite{WelinderEtal2010}}
\newcommand{\abof}{\acf{ABO}~\cite{collins2022abo}}

\AtBeginDocument{%
  \providecommand\BibTeX{{%
    \normalfont B\kern-0.5em{\scshape i\kern-0.25em b}\kern-0.8em\TeX}}}

\acrodef{CtI}{category-to-image}

\author{
Mariya Hendriksen\inst{1}\orcidID{0000-0003-0314-2955}
\and
Svitlana Vakulenko\inst{2}\orcidID{0000-0002-5278-8886}\thanks{Research conducted while the author was at the University of Amsterdam.}
\and
Ernst Kuiper\inst{3}\orcidID{0000-0002-8075-4894}
\and
Maarten de Rijke\inst{4}\orcidID{0000-0002-1086-0202}
\institute{
AIRLab,  University of Amsterdam, The Netherlands\\
\email{m.hendriksen@uva.nl}
\and
Amazon, Spain\\
\email{svvakul@amazon.com}
\and
Bol.com, The Netherlands\\
\email{ekuiper@bol.com}
\and
University of Amsterdam, The Netherlands\\
\email{m.derijke@uva.nl}
}
}

\author{
Mariya Hendriksen\inst{1}
\and
Svitlana Vakulenko\inst{2}\thanks{Research conducted while the author was at the University of Amsterdam.}
\and
Ernst Kuiper\inst{3}
\and
Maarten de Rijke\inst{4}
\institute{
AIRLab,  University of Amsterdam, The Netherlands\\
\email{m.hendriksen@uva.nl}
\and
Amazon, Spain\\
\email{svvakul@amazon.com}
\and
Bol.com, The Netherlands\\
\email{ekuiper@bol.com}
\and
University of Amsterdam, The Netherlands\\
\email{m.derijke@uva.nl}
}
}

\authorrunning{M.  Hendriksen et al.}

\newcommand\sD{\ensuremath{\mathcal{D}}}

\newcommand\sI{\ensuremath{\mathcal{I}}}

\newcommand\sT{\ensuremath{\mathcal{T}}}

\newcommand\bq{\ensuremath{\mathbf{q}}}

\newcommand\bx{\ensuremath{\mathbf{x}}}

\usepackage{color}

\begin{document}

\title{Scene-centric vs.\ Object-centric Image-Text \OURTASK{}: A Reproducibility Study}
\titlerunning{Scene-centric vs.\  Object-centric \OURTASK{}}

\maketitle 

\begin{abstract}
Most approaches to \ac{CMR} focus either on ob\-ject-centric datasets, meaning that each document depicts or describes a single object, or on scene-centric datasets, meaning that each image depicts or describes a complex scene that involves multiple objects and relations between them.
We posit that a robust \ac{CMR} model should generalize well across both dataset types. 
Despite recent advances in \ac{CMR}, the reproducibility of the results and their generalizability across different dataset types has not been studied before.
We address this gap and focus on the reproducibility of the \acl{SOTA} \ac{CMR} results when evaluated on object-centric and scene-centric datasets.  We select two \acl{SOTA} \ac{CMR} models with different architectures: 
\begin{enumerate*}[label=(\roman*)]
\item \acs{CLIP};  and 
\item X-VLM.
\end{enumerate*}
Additionally, we select two scene-centric datasets, and three object-centric datasets, and determine the relative performance of the selected models on these datasets. 
We focus on reproducibility, replicability, and generalizability of the outcomes of previously published \ac{CMR} experiments. We discover that the experiments are not fully reproducible and replicable. Besides, the relative performance results partially generalize across object-centric and scene-centric datasets. On top of that, the scores obtained on object-centric datasets are much lower than the scores obtained on scene-centric datasets. 
For reproducibility and transparency we make our source code and the trained models publicly available.

\end{abstract}

\acresetall


\section{Introduction}
\label{sec:introduction}

\Ac{CMR} is the task of finding relevant items across different modalities.
For example, given an image, find a text or vice versa.
The main challenge in \ac{CMR} is known as \emph{the heterogeneity gap}~\citep{,carvalho2018cross, hu2019scalable}.
Since items from different modalities have different data types, the similarity between them cannot be measured directly.
Therefore, the majority of \ac{CMR} methods published to date attempt to bridge this gap by learning a latent representation space, where the similarity between items from different modalities can be measured~\cite{wang2016comprehensive}. 

In this work, we specifically focus on \emph{image-text} \ac{CMR}, which uses textual and visual data. The retrieval task is performed on \emph{image-text pairs}. 
In each image-text pair, the text (often referred to as \emph{caption}) describes the corresponding image it is aligned with. 
For image-text \ac{CMR} we use either an image or a text as a query~\citep{wang2016comprehensive}.
Hence, the \ac{CMR} task that we address in this paper consists of two subtasks:
\begin{enumerate*}[label=(\roman*)]
	\item \emph{text-to-image retrieval}: given a text that describes an image, retrieve all the images that match this description; and
	\item \emph{image-to-text retrieval}: given an image, retrieve all texts that can be used to describe this image.
\end{enumerate*}

\header{Scene-centric vs.\ object-centric datasets} 
Existing image datasets can be grouped into \emph{scene-centric} and \emph{object-centric} datasets~\citep{zhang2021mosaicos, shen2019deep}.
The two types of datasets are typically used for different tasks, viz.\ the tasks of scene and object understanding, respectively.
They differ in important ways that are of interest to us when evaluating performance and generalization abilities of \ac{CMR} models.

Scene-centric images depict complex scenes that typically feature multiple objects and relations between them.  
These datasets contain image-text pairs, where, in each pair, an image depicts a complex scene of objects and the corresponding text describes the whole scene, often focusing on \textit{relations and activities}.

Images in object-centric image datasets are usually focused on a single object of interest that they primarily depict.
This object is often positioned close to the center of an image with other objects, optionally, in the background. 
Object-centric datasets contain image-text pairs, where, in each pair, an image depicts an object of interest and the corresponding text describes the depicted \textit{object and its (fine-grained) attributes}. 

\begin{figure*}[t]
\centering
   \includegraphics[height=5.35cm]{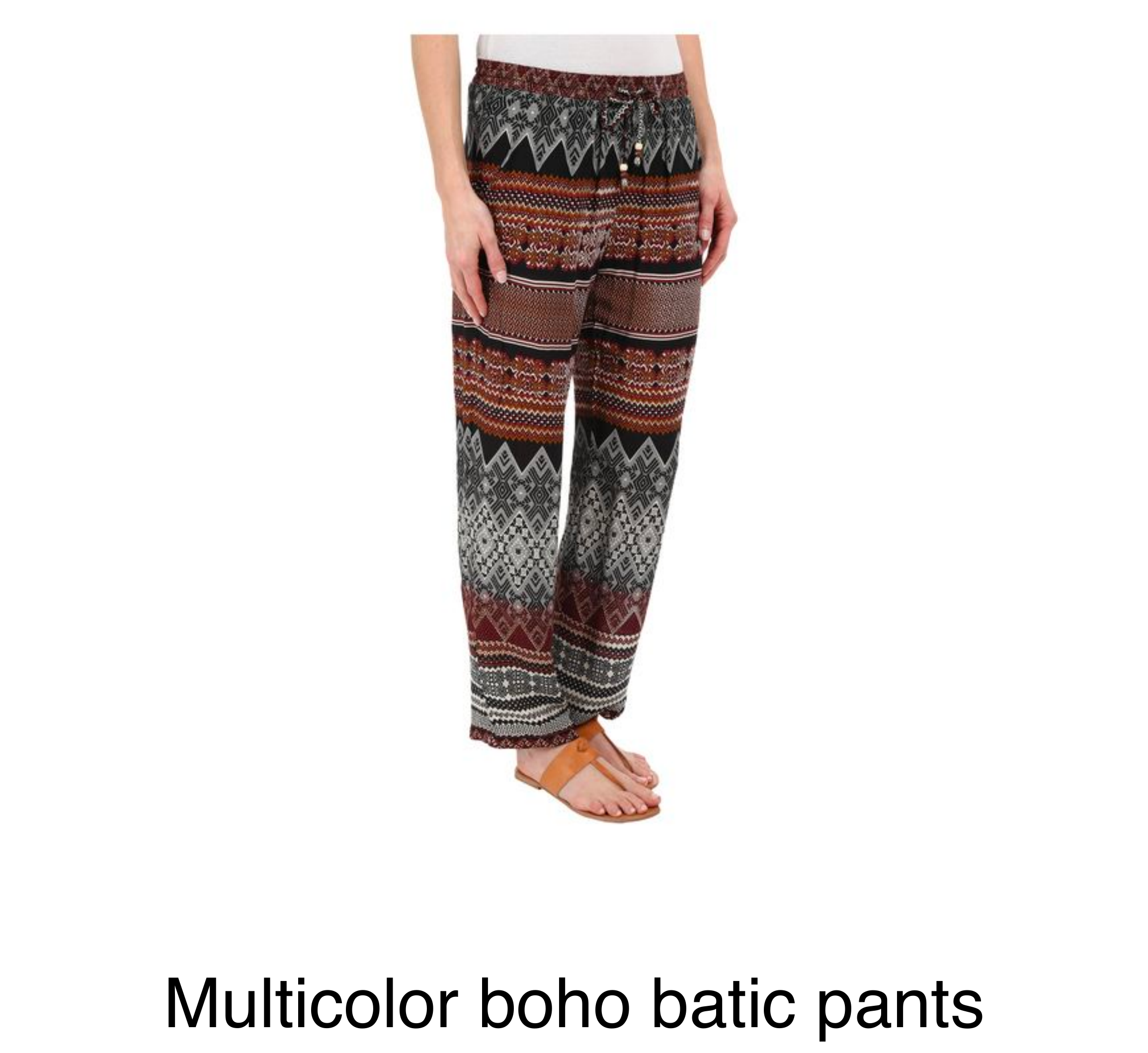}
   \includegraphics[height=5.35cm]{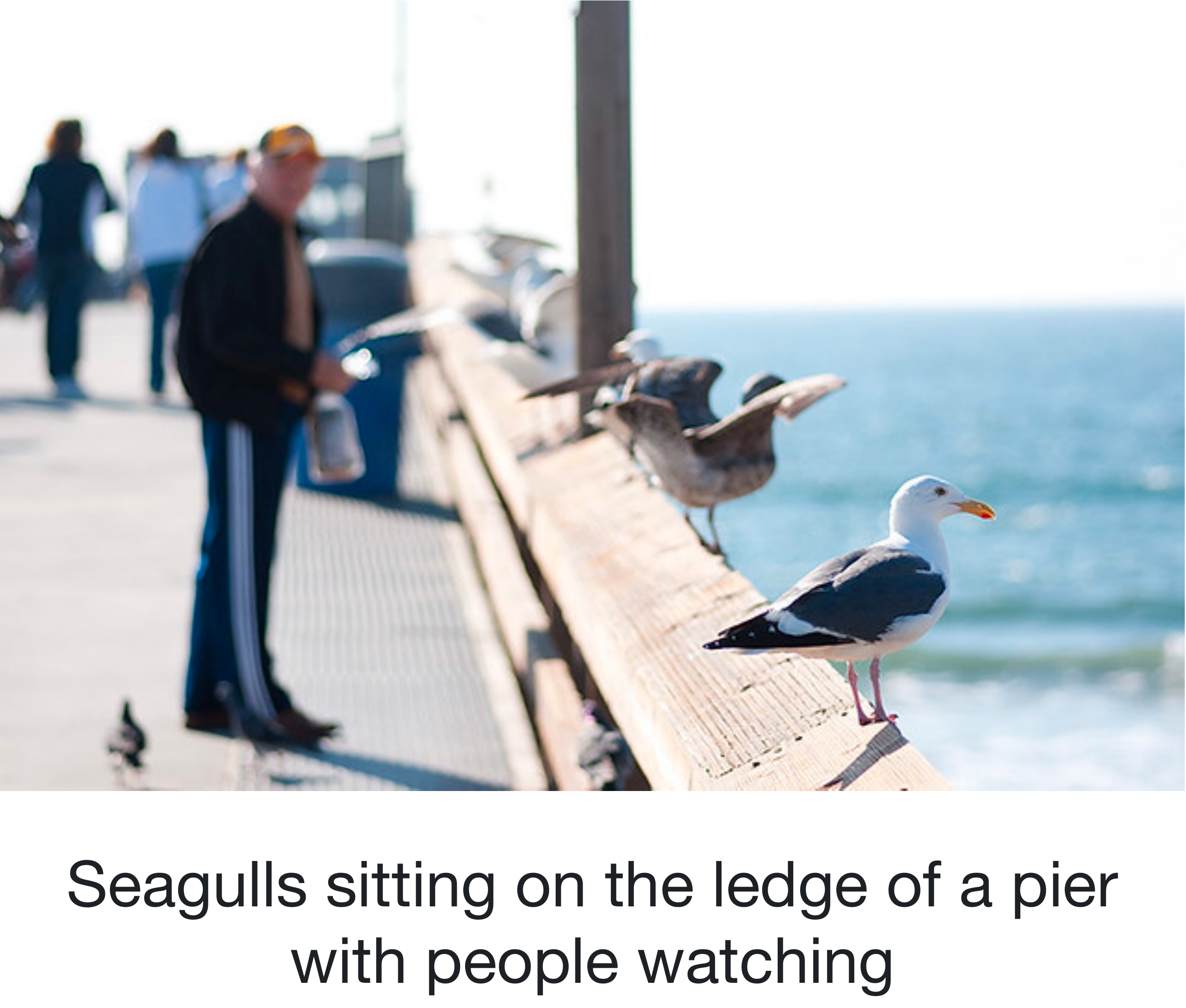}
   \caption{An object-centric (left) and a scene-centric (right) image-text pair. Sources: Fashion200k (left); MS COCO (right).
}
 \label{fig:object-centric-vs-scene-centric}
\end{figure*}

To illustrate the differences between the two dataset types in \ac{CMR}, we consider the examples provided in Fig.~\ref{fig:object-centric-vs-scene-centric} with an object-centric \ic{} (left) and a scene-centric \ic{} (right). 
Note how the pairs differ considerably in terms of the visual style and the content of the caption.
The pair on the left focuses on a single object (``pants'') and describes its fine-grained visual attributes (``multicolor,'' ``boho,'' ``batic'').
The pair on the right captures a scene describing multiple objects (``seagulls,'' ``pier,'' ``people'') and relations between them (``sitting,'' ``watching''). 

\header{Research goals}
We focus on (traditional) \ac{CMR} methods that extract features from each modality and learn a common representation space.
Recent years have seen extensive experimentation with such \ac{CMR} methods, mostly organized into two groups: 
\begin{enumerate*}[label=(\roman*)]
	\item contrastive experiments on object-centric datasets~\citep{han2017automatic}, and
	\item contrastive experiments on scene-centric datasets~\citep{lin2014microsoft}.
\end{enumerate*}
In this paper, we consider representative \acl{SOTA} \ac{CMR} methods from both groups. We select two pre-trained models which demonstrate \acl{SOTA} performance on \ac{CMR} task and evaluate them in a zero-shot setting. 
In line with designs used in prior reproducibility work on \ac{CMR}~\cite{bleeker-2022-do} we select two models for the study.
Following the ACM terminology \cite{acm_badge}, we focus on \emph{reproducibility} (different team, same experimental setup) and \emph{replicability} (different team, different experimental setup) of previously reported results. And following \citet{voorhees-2002-philosophy}, we focus on relative (a.k.a.\ comparative) performance results. In addition, for the reproducibility experiment, we consider the absolute difference between the reported scores and the reproduced scores.

We address the following research questions:
\begin{enumerate*}[label=(RQ\arabic*),leftmargin=*,nosep]
	\item Are published relative performance results on \ac{CMR} reproducible?
This question matters because it allows us to confirm the validity of reported results. We show that the relative performance results are not fully reproducible. Specifically, the results are reproducible for one dataset, but not for the other dataset). 
\end{enumerate*}

We then shift to replicability and examine whether lessons learned on scene-centric datasets transfer to object-centric datasets:
\begin{enumerate*}[label=(RQ\arabic*),leftmargin=*,nosep,resume]
	\item To what extent are the published relative performance results replicable?
That is, we investigate the validity of the reported results when evaluated in a different setup.
We find that relative performance results are partially replicable, using other datasets. 
\end{enumerate*}

After investigating the reproducibility and replicability of the results, we consider the generalizability of the results. We contrastively evaluate the results on object-centric and scene-centric datasets:
\begin{enumerate*}[label=(RQ\arabic*),leftmargin=*,nosep,resume]
	\item Do relative performance results for \acl{SOTA} \ac{CMR} methods generalize from scene-centric datasets to object-centric datasets?
We discover that the relative performance results only partially generalize across the two dataset types.
\end{enumerate*}

\header{Main contributions}
Our main contributions are: 
\begin{enumerate*}[label=(\roman*)]
	\item We are one of the first to consider reproducibility in the context of \ac{CMR} and reproduce scene-centric \ac{CMR} experiments from two papers~\citep{radford2021learning, DBLP:conf/icml/ZengZL22} and find that the results are only partially reproducible.
	\item We perform a replicability study and examine whether relative performance differences reported for \ac{CMR} methods generalize from scene-centric datasets to object-centric datasets.
	\item We investigate the generalizability of obtained results and analyze the effectiveness of pre-training on scene-centric datasets for improving the performance of \ac{CMR} on object-centric datasets, and vice versa. And, finally,
	\item to facilitate the reproducibility of our work, we provide the code and the pre-trained models used in our experiments.\footnote{\url{https://github.com/mariyahendriksen/ecir23-object-centric-vs-scene-centric-CMR}}
\end{enumerate*}

\section{Related Work}
\label{sec:related_work}

\textbf{\Ourtask{}.} 
\ac{CMR} methods attempt to construct a multimodal representation space, where the similarity of concepts from different modalities can be measured. 
Some of the earliest approaches in \ac{CMR} utilised canonical correlation analysis~\cite{gong2014improving,klein2014fisher}. 
They were followed by a dual encoder architecture equipped with a recurrent and a convolutional component, a hinge loss~\citep{frome2013devise, wang2016learning} and hard-negative mining~\cite{faghri2017vsepp}. 
Later on, several attention-based architectures were introduced such as architectures with dual attention~\cite{nam2017dual}, stacked cross-attention~\cite{lee2018stacked}, bidirectional focal attention~\cite{liu2019focus}. 

Another line of work proposed to use transformer encoders~\cite{vaswani2017attention} for \ac{CMR} task~\cite{messina2021fine}, and adapted the BERT model~\cite{devlin2018bert} as a backbone~\citep{gao2020fashionbert, zhuge2021kaleido}.
Some other researchers worked on improving \ac{CMR} via modality-specific graphs~\cite{wang2021cross}, or image and text generation modules~\cite{gu2018look}. 

There is also more domain-specific work that focused on \ac{CMR} in fashion~\citep{laenen2018web, laenen2017cross, goei2021tackling, laenen2022cross}, e-commerce~\citep{hendriksen2022extending, hendriksen2022multimodal}, cultural heritage~\cite{sheng2021fine} and cooking~\cite{wang2021cross}.

In contrast to the majority of prior work on the topic, we focus on the reproducibility, replicability, and generalizability of \ac{CMR} methods. In particular, we explore the \acl{SOTA} models designed for the \ac{CMR} task by examining their performance on scene-centric and object-centric datasets.

\header{Scene-centric and object-centric datasets} 
The majority of prior work related to object-centric and scene-centric datasets focuses on \acl{CV} tasks such as object recognition, object classification, and scene recognition.
\citet{herranz2016scene} investigated biases in a \acs{CNN} when trained on scene-centric versus object-centric datasets and evaluated on the task of object classification.

In the context of object detection, prior work focused on combining feature representations learned from object-centric and scene-centric datasets to improve the performance when detecting small objects~\cite{shen2019deep}, and using object-centric images to improve the detection of objects that do not appear frequently in complex scenes~\cite{zhang2021mosaicos}.
Finally, for the task of scene recognition, \citet{zhou2014learning} explored the quality of feature representations learned from both scene-centric and object-centric datasets and applied them to the task of scene recognition.

Unlike prior work on the topic, in this paper, we focus on both scene-centric and object-centric datasets for evaluation on \ac{CMR} task. In particular, we explore how \ac{SOTA} \ac{CMR} models perform on object-centric and scene-centric datasets.

\header{Reproducibility in \ourtask{}}
To the best of our knowledge, despite the popularity of the \ac{CMR} task, there are very few papers that focus on reproducibility of research in \ac{CMR}. 
Some rare (recent) examples include \citep{bleeker-2022-do}, where the authors survey metric learning losses used in \acl{CV} and explore their applicability for \ac{CMR}. 
\citet{DBLP:conf/sigir/RaoW0QZLT22} analyze contributing factors that affect the performance of the \acl{SOTA} \ac{CMR} models. 
However, all prior work focuses on exploring model performance only on two popular scene-centric datasets: \coco{} and \flickr{}. 

In contrast, in this work, we take advantage of the diversity of the \ac{CMR} datasets and specifically focus on examining how the \acl{SOTA} \ac{CMR} models perform across different dataset types: scene-centric and object-centric datasets.


\section{Task Definition}
\label{sec:task_definition}

We follow the same notation as in previous work~\citep{zhang2020contrastive, varamesh2020self, brown2020smooth}. 
An image-caption cross-modal dataset consists of a set of images $\sI$ and texts $\sT$ where the images and texts are aligned as image-text pairs: $\sD{} = \{ (\bx_{\sI}^1$, $\bx_{\sT}^1)$, \ldots, $(\bx_{\sI}^n, \bx_{\sT}^n) \}$.

The \acfi{CMR} task is defined analogous to the standard \acl{IR} task: given a query $\bq$ and a set of $m$ candidates $\Omega_{\bq} = \{ \bx^1, \dots, \bx^m \}$ we aim to rank all the candidates w.r.t.\ their relevance to the query $\bq$.
In \ac{CMR}, the query can be either a text $\bq_{\sT}$ or an image $\bq_{\sI}$: $\bq \in \{ \bq_{\sT}, \bq_{\sI}\}$.
Similarly,  the set of candidate items can be either visual $\sI_{\bq} \subset \sI$,  or textual $\sT_{\bq} \subset \sT $ data: $\Omega \in \{ \sI_{\bq},  \sT_{\bq} \}$.

The \ac{CMR} task is performed across modalities,  therefore, if the query is a text then the set of candidates are images, and vice versa. 
Hence, the task comprises effectively two subtasks:
\begin{enumerate*}[label=(\roman*)]
	\item \emph{text-to-image retrieval}: given a textual query $\bq_{\sT}$ and a set of candidate images $\Omega \subset \sI$,  we aim to rank all instances in the set of candidate items $\Omega$ w.r.t.\ their relevance to the query $\bq_{\sT}$;
	\item \emph{image-to-text retrieval}: given an image as a query $\bq_{\sI}$ and a set of candidate texts $\Omega \subset \sT$,  we aim to rank all instances in the set of candidate items $\Omega$ w.r.t.\ their relevance to the query $\bq_{\sI}$. 
\end{enumerate*}


\section{Methods}
\label{sec:approach}
In this section, we give an overview of the models included in the study, of the models which were excluded, and provide justification for it.
All the approaches we focus on belong to the traditional \ac{CMR} framework and comprise two stages.  First, we extract textual and visual features.  The features are typically extracted with a textual encoder and a visual encoder.  Next, we learn a latent representation space where the similarity of items from different modalities can be measured directly.

\subsection{Methods included for comparison}
We focus on \ac{CMR} in \emph{zero-shot setting}, hence, we only consider pre-trained models. Therefore, we focus on the models that are released for public use. Besides, as explained in Section~\ref{sec:introduction}, we follow prior reproducibility work to inform our experimental choices regarding the number of models. 
Given the above-mentioned requirements, we selected two methods that demonstrate \acl{SOTA} performance on the \ac{CMR} task: CLIP and X-VLM.

\header{\ac{CLIP}~\cite{radford2021learning}}
This model is a dual encoder that comprises an image encoder, and a text encoder. The model was pre-trained in a contrastive manner using a symmetric loss function. It is trained on 400 million image-caption pairs scraped from the internet.
The text encoder is a transformer~\cite{vaswani2017attention} with modification from~\cite{radford2019language}.
For the image encoder, the authors present two architectures. The first one is based on ResNet~\cite{he2016deep} and it is represented in five variants in total. The first two options are ResNet-50, ResNet-101; the last three options are variants of ResNet scaled up in the style of EfficientNet~\cite{tan2019efficientnet}
The second image encoder architecture is a \ac{ViT}~\cite{dosovitskiy2020image}. It is presented in three variants: ViT-B/32, a ViT-B/16, and a ViT-L/14.
The \ac{CMR} results reported in the original paper are obtained with a model configuration where vision transformer ViT-L/14 is used as an image encoder, and the text transformer is a text encoder. Hence, we use this configuration in our experiments.

\header{X-VLM~\cite{DBLP:conf/icml/ZengZL22}} This model consists of three encoders: an image encoder, a text encoder, and a cross-modal encoder.
The image and text encoder take an image and text as inputs and output their visual and textual representations. The cross-modal encoder fuses the output of the image encoder and the output of the text encoder. The fusion is done via a cross-attention mechanism. For \ac{CMR} task, the model is fine-tuned via a contrastive learning loss and a matching loss. 
All encoders are transformer-based.
The image encoder is a \ac{ViT} initialised with Swin Transformer$_{base}$~\cite{liu2021swin}.
Both the text encoder and the cross-modal encoder are initialised using different layers of BERT~\cite{devlin2018bert}: the text encoder is initialized using the first six layers, whereas the cross-modal encoder is initialised using the last six layers.

\subsection{Methods excluded from comparison} 
While selecting the models for the experiments, we considered other architectures with promising performance on the MS COCO and the Flickr30k datasets. Below, we outline the architectures we considered and explain why they were not included.

Several models such as Visual N-Grams~\cite{li2017learning}, Unicoder-VL~\cite{li2020unicoder}, ViLT-B/32~\cite{kim2021vilt}, UNITER~\cite{chen2019uniter} were excluded because they were consistently outperformed by CLIP on the MS COCO and Flickr30k datasets by large margins. 
Besides, we excluded ImageBERT~\cite{qi2020imagebert} because it was outperformed by CLIP on the MS COCO dataset.
ALIGN~\cite{jia2021scaling}, ALBEF~\cite{li2021align}, VinVL~\cite{zhang2021vinvl}, METER~\cite{dou2022empirical} were not included because X-VLM consistently outperformed them.
UNITER~\cite{chen2019uniter} was beaten by both CLIP and X-VLM.
We did not include other well-performing models such as ALIGN~\cite{jia2021scaling}, Flamin\-go~\cite{alayrac2022flamingo}, CoCa~\cite{yu2022coca} because the pre-trained models were not publicly available.


\section{Experimental Setup}
\label{sec:experiments}

In this section, we discuss our experimental design including the choice of datasets, subtasks, metrics, and implementation details.

\subsection{Datasets} 
We run experiments on two scene-centric and three object-centric datasets.  Below, we discuss each of the datasets in more detail.

\header{Scene-centric datasets} We experiment with two scene-centric datasets:
\begin{enumerate*}[label=(\roman*)]
\item \cocof{} contains  \numprint{123287} images depicting regular scenes from everyday life with multiple objects placed in their natural contexts. There are 91 different object types such as ``person'', ``bicycle'', ``apple''.
\item \flickr{} contains \numprint{31783} images of regular scenes from everyday life, activities, and events. 
\end{enumerate*}
For both scene-centric datasets, we use the splits provided in~\citep{karpathy2015deep}. The MS COCO dataset is split into \numprint{113287} images for training, \numprint{5000} for testing and \numprint{5000} for validation; the Flickr30k dataset has \numprint{29783} images for training, \numprint{1000} for testing and \numprint{1000} for validation. In both datasets, every image was annotated with five captions using Amazon Mechanical Turk. Besides, we select one caption per image randomly and use the test set for our experiments.

\header{Object-centric datasets} We consider three object-centric datasets in our experiments:
\begin{enumerate*}[label=(\roman*)]
\item \cubf{} contains \numprint{11788} images of 200 birds species. Each image is annotated with a fine-grained caption from~\cite{reed2016learning}. We selected one caption per image randomly. Each caption is at least 10 words long and does not contain any information about the birds' species or actions. 

\item \fashion{} contains \numprint{209544} images that depict various fashion items in five product categories (dress, top, pant, skirt, jacket) and their corresponding descriptions.

\item \abof{} contains \numprint{147702} product listings associated with \numprint{398212} images. This dataset was derived from Amazon.com product listings. We selected one image per listing and used the associated product description as its caption.
The majority of images depict a single product on a white background. The product is located in the center of the image and takes at least 85\% of the image area.
\end{enumerate*}
For all object-centric datasets, we use the splits provided by the dataset authors and use the test split for our experiments.

\subsection{Subtasks} 
\begin{figure*}[t]
\centering
   \includegraphics[width=\textwidth]{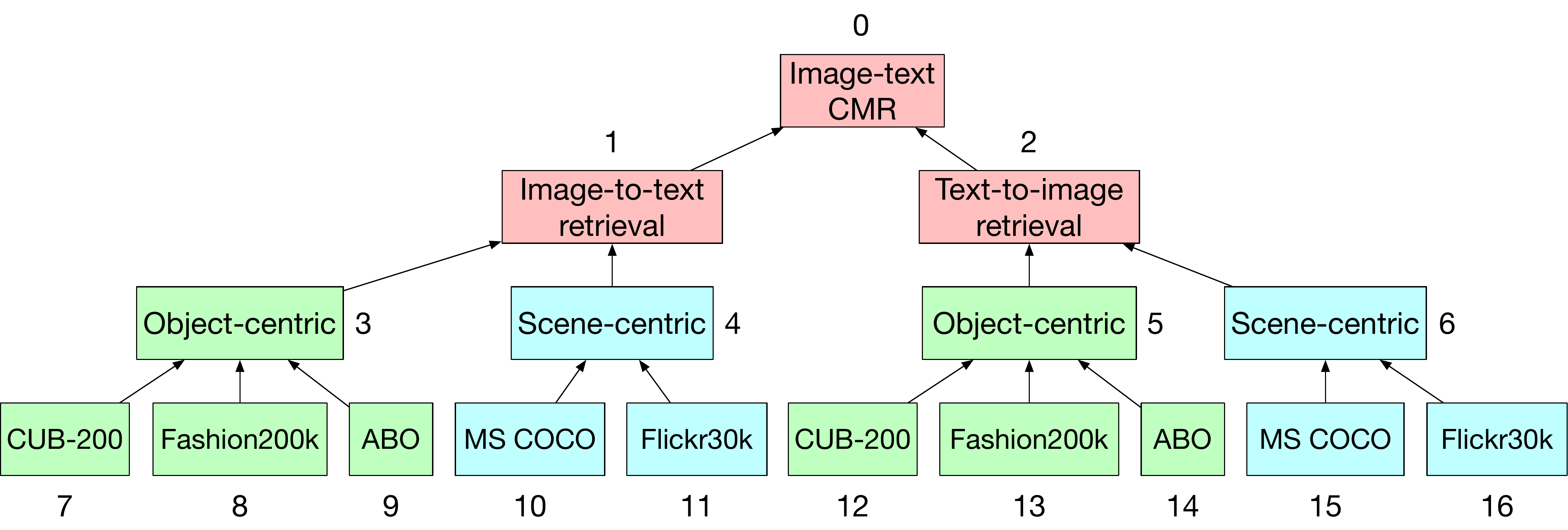}
   \vspace*{1mm}
   \caption{Our experimental design for evaluating \ac{CMR} methods across object-centric and scene-centric datasets. The blue colour indicates parts of the tree used in Experiment~1, the green color indicates parts of the tree used in Experiment~2, and the red color indicates parts used in all experiments. (Best viewed in color.)}
 \label{fig:framework-overview}
\end{figure*}

Our goal is to assess and compare the performance of the \ac{CMR} methods (described in Section~\ref{sec:experiments}) across the object-centric and scene-centric datasets described in the previous subsection.
We design an experimental setup that takes into account two \ac{CMR} subtasks and two dataset types.
It can be summarized using a tree with branches that correspond to different configurations (see Fig.~\ref{fig:framework-overview}).
We explain how we cover the branches of this tree in the next subsection.

The tree starts with a root (``Image-text \ac{CMR}'' with label $0$) that has sixteen descendants, in total. The root node has two children corresponding to the two image-text \ac{CMR} subtasks: text-to-image retrieval (node 1) and image-to-text retrieval (node 2). Since we want to evaluate each of these subtasks on both object-centric and scene-centric datasets, nodes 1 and 2 also have two children each, i.e., the nodes $\{3, 4, 5, 6 \}$.
Finally, every object-centric node has three children: CUB-200, Fashion200k, and ABO datasets $\{7$, $8$, $9$, $12$, $13$, $14 \}$; and every scene-centric node has two children: MS COCO and Flickr30k datasets $\{10, 11, 15, 16 \}$. 

\subsection{Experiments}

To answer the research questions introduced in Section~\ref{sec:introduction},  we conduct two experiments.
In all the experiments, we use CLIP and X-VLM models in a zero-shot setting. Following \cite{voorhees-2002-philosophy}, we focus on relative performance results.
In each experiment, we consider different subtrees from Fig.~\ref{fig:framework-overview}.
Following \citep{radford2021learning, DBLP:conf/icml/ZengZL22, li2017learning, li2020unicoder, kim2021vilt}, we use Recall@K where $K = \{ 1, 5, 10 \}$ to evaluate the model performance in all our experiments.
In addition, following \citep{ueki2021survey, zhang2022negative, song2021image}, we calculate the sum of recalls (rsum) for text-to-image, and image-to-text retrieval tasks as well as the total sum of recalls for both tasks.

For text-to-image retrieval, we first obtain representations for all the candidate images by passing them through the image encoder of the model.
Then we pass each textual query through the text encoder of the model and retrieve the top-$k$ candidates ranked by cosine similarity w.r.t. the query.

For image-to-text retrieval, we do the reverse, using the texts as candidates and images as queries.
More specifically, we start by obtaining representations of the candidate captions by passing them through the text encoder.
Afterwards, for each of the visual queries, we pass the query through the image encoder and retrieve top-$k$ candidates ranked by cosine similarity w.r.t. the query. 

In \emph{Experiment 1} we evaluate the reproducibility of the \ac{CMR} results reported in the original publications (RQ1).
Both models we consider (CLIP and X-VLM) were originally evaluated on two scene-centric datasets, viz.\ \coco{} and \flickr{}. Therefore, for our reproducibility study, we also evaluate these models on these two datasets. We evaluate both text-to-image and image-to-text retrieval. That is, we focus on the two sub-trees 0$\leftarrow$1$\leftarrow$4$\leftarrow$\{10, 11\} and 0$\leftarrow$2$\leftarrow$6$\leftarrow$\{15, 16\} (the red and blue parts of the tree) from Fig.~\ref{fig:framework-overview}.
In addition to relative performance results, we consider absolute differences between the reported scores and the reproduced scores.
Following \citet{petrov2022systematic}, we assume that the score is reproduced if we obtain a score value equal to the reported score given a relative tolerance of $\pm5\%$. 

In \emph{Experiment 2} we focus on the replicability of the reported results on object-centric datasets (RQ2). Thus, we evaluate CLIP and X-VLM on the \cub{}, \fashion{}, and \abo{} datasets. This experiment covers the subtrees 0$\leftarrow$1$\leftarrow$3$\leftarrow$\{7, 8, 9\} and 0$\leftarrow$2 $\leftarrow$5$\leftarrow$\{12, 13, 14\} (the red and green parts of the tree) in Fig.~\ref{fig:framework-overview}.

After obtaining the results from Experiment~1 and~2, we examine the generalizability of the obtained scores (RQ3). We do so by comparing the relative performance results the models achieve on the object-centric versus scene-centric datasets. More specifically, we compare the relative performance of CLIP and X-VLM  on \cub{}, \fashion{}, \abo{} with their relative performance on \coco{} and \flickr{}. Thus, this experiment captures the complete tree in Fig.~\ref{fig:framework-overview}.

\section{Results}
\label{sec:results}

We focus on the reproducibility (different team, same setup) and replicability (different team, different setup) of the \ac{CMR} experiments reported in the original papers devoted to CLIP~\cite{radford2021learning} and  X-VLM~\cite{DBLP:conf/icml/ZengZL22}. 
To organize our result presentation, we refer to the tree in Fig.~\ref{fig:framework-overview}. We traverse the tree bottom up, from the leaves to the root. 

\subsection{RQ1: Reproducibility} 
To address RQ1, we report on the outcomes of Experiment~1. 
We investigate to what extent the \ac{CMR} results reported in the original papers devoted to CLIP~\cite{radford2021learning} and X-VLM~\cite{DBLP:conf/icml/ZengZL22} are reproducible. Given that both methods were originally evaluated on two scene-centric datasets, viz.\ \coco{} and \flickr{}, we evaluate the models on the text-to-image and image-to-text tasks on these two datasets. Therefore, we focus on the two blue sub-trees 0$\leftarrow$1$\leftarrow$4$\leftarrow$\{10, 11\} and 0$\leftarrow$2$\leftarrow$6$\leftarrow$\{15, 16\} from Fig.~\ref{fig:framework-overview}.

\begin{table}[t]
\centering
\caption{Results of Experiment 1 (reproducibility study), using the MS COCO and Flickr30k datasets. ``Orig.'' indicates the scores from the original publications. ``Repr.'' indicates the scores that we obtained.
}
\label{tab:repro}
\begin{tabular}{ll ccc c ccc c ccc}
\toprule
 &  & \multicolumn{3}{c}{\textbf{Text-to-image}} && \multicolumn{3}{c}{\textbf{Image-to-text}} && \multicolumn{3}{c}{\textbf{Rsum}} \\
\cmidrule{3-5}\cmidrule{7-9}\cmidrule{11-13}
 & \textbf{Model} & 
 \multicolumn{1}{c}{\textbf{R@1}} & \multicolumn{1}{c}{\textbf{R@5}} & \multicolumn{1}{c}{\textbf{R@10}} && 
\multicolumn{1}{c}{\textbf{R@1}} & \multicolumn{1}{c}{\textbf{R@5}} & \multicolumn{1}{c}{\textbf{R@10}} && 
\multicolumn{1}{c}{\textbf{t2i}} & \multicolumn{1}{c}{\textbf{i2t}} & \multicolumn{1}{c}{\textbf{total}} \\ 
\midrule
 &  & \multicolumn{9}{c}{\textbf{MS COCO (5k)}} \\ 
\midrule
\multirow{2}{*}{\rotatebox{90}{Orig.}}
 & CLIP~\cite{radford2021learning} & 
 37.80 & 62.40 & 72.20 &&
 58.40 & 81.50 & 88.10 && 
 172.40 & 228.00 & 400.40 
\\
 & X-VLM~\cite{DBLP:conf/icml/ZengZL22} & 
 \textbf{55.60} & \textbf{82.70} & \textbf{90.00} && 
 \textbf{70.80} & \textbf{92.10} & \textbf{96.50} && 
 \textbf{228.30} & \textbf{259.40} & \textbf{487.70} \\
 \midrule
\multirow{2}{*}{\rotatebox{90}{Repr.}}
& CLIP & 
 21.59 & 40.22 & 49.80 && 
 24.36 & 44.13 & 53.41 && 
 111.61 & 121.90 & 233.51 
\\
& X-VLM & 
\textbf{42.79} & \textbf{67.61} & \textbf{67.64} &&
 \textbf{64.60} & \textbf{84.48} & \textbf{84.50} &&
  \textbf{178.04} & \textbf{233.58} & \textbf{411.62}
\\
\midrule
 &  & \multicolumn{9}{c}{\textbf{Flickr30k (1k) }} \\
\midrule
\multirow{2}{*}{\rotatebox{90}{Orig.}}
& CLIP~\cite{radford2021learning} & 
68.70 & 90.60 & 95.20 && 
\textbf{88.00} & \textbf{98.70} & 99.40 && 
254.50 & \textbf{286.10} & 540.60 
\\
& X-VLM~\cite{DBLP:conf/icml/ZengZL22} & 
\textbf{71.90} & \textbf{93.30} & \textbf{96.40} && 
85.30 & 97.80 & \textbf{99.60} && 
\textbf{261.60} & 282.70 & \textbf{544.30} 
\\
\midrule
\multirow{2}{*}{\rotatebox{90}{Repr.}}
& CLIP & 
\textbf{74.95} & \textbf{93.09} & \textbf{96.15} && 
\textbf{77.02} & \textbf{94.18} & \textbf{96.84} && 
\textbf{264.19} & \textbf{268.04} & \textbf{532.23} 
\\
& X-VLM & 
37.82 & 82.36 & 82.48 &&
63.30 & 91.10 & 91.10 &&
202.66 & 245.50 & 448.16 
\\ 
\bottomrule
\end{tabular}
\end{table}

\header{Results}
The results of Experiment~1 are shown in Table~\ref{tab:repro}.
We recall the scores obtained in the original papers~\citep{radford2021learning,DBLP:conf/icml/ZengZL22} (``Orig.'') and the scores that we obtained (``Repr.''), on the MS COCO and Flickr30k datasets.
Across the board, the scores that we obtained (the ``reproduced scores'') tend to be lower than the scores obtained in the original publications (the ``original scores'').

On the MS COCO dataset, X-VLM consistently outperforms CLIP, both in the original publications and in our setup, for both the text-to-image and the image-to-text tasks. Moreover, this holds for all R@$n$ metrics, and, hence, for the Rsum metrics.
Interestingly, the relative gains that we obtain tend to be larger than the ones obtained in the original publications. For example, our biggest relative difference is for the image-to-text task in terms of the R@1 metric: according to the scores reported in \citep{DBLP:conf/icml/ZengZL22, radford2021learning}, X-VLM outperforms CLIP by 21\%, whereas in our experiments the relative gain is 165\%.

On average, the original CLIP scores are as much as $\sim$70\% higher than the reproduced scores; the original scores for X-VLM are $\sim$20\% higher than the reproduced ones.
When considering the absolute differences between the original scores and the reproduced scores and assuming a relative tolerance of ±5\%, we see that, on the MS COCO dataset, the scores are not reproducible for both models. 

On the Flickr30k dataset, we see a different pattern. For the text-to-image task, the original results indicate that X-VLM consistently outperforms CLIP, on all R@$n$ metrics, but according to our results, the relative order is consistently reversed. For the image-to-text task, we obtained mixed outcomes: for R@1 and R@5, the original order (CLIP outperforms X-VLM) is confirmed, but for R@10 the order is swapped. According to our experimental results, however, CLIP consistently outperforms X-VLM on all tasks, and on all R@$n$ metrics (and hence also on the Rsum metrics).

On the Flickr30k dataset, the CLIP scores are reproduced on the text-to-image and image-to-text retrieval tasks when the model is evaluated on R@5 and R@10.
On the text-to-image task, the reproduced R@5 score is 2.7\% higher than the original score; the reproduced R@10 score is 1\% higher than the original score.
For the image-to-text retrieval task, the reproduced R@5 score is 4\% lower than the original score; the reproduced R@10 score is 2\% lower than the original score.

\header{Answer to RQ1}
In the case of the CLIP model, the obtained \emph{absolute} scores were reproducible only on the Flickr30k dataset for the text-to-image and the image-to-text tasks when evaluated on R@5 and R@10.
For X-VLM, we did not find the absolute scores obtained when evaluating the model on the MS COCO and Flickr20k datasets to be reproducible, neither for the text-to-image nor the image-to-text tasks.

The \emph{relative} outcomes on the MS COCO dataset could be reproduced, for all tasks and metrics, whereas on the Flickr30k dataset, they could only partially be reproduced, that is, only for the image-to-text task on the R@1 and R@5 metrics; for the text-to-image task, X-VLM outperforms CLIP according to the original scores, but CLIP outperforms X-VLM according to our reproduced scores. 

\header{Upshot} 
As explained in Section~\ref{sec:approach}, in this paper we focus on \ac{CMR} in a zero-shot setting. 
This implies that the differences that we observed between the original scores and the reproduced scores must be due to differences in text and image data (pre-)processing and loading.
We, therefore, recommend that the future work includes (as much as is practically possible) tools and scripts used in these stages of the experiment with the publication of its implementations.

\subsection{RQ2: Replicability} 
To answer RQ2, we replicate the originally reported text-to-image and image-to-text retrieval experiments in a different setup, i.e., by evaluating CLIP and X-VLM using object-centric datasets instead of scene-centric datasets.  Thus, we evaluate CLIP and X-VLM on the \cub{}, \fashion{}, and \abo{} datasets and focus on the green subtrees 0$\leftarrow$1$\leftarrow$3$\leftarrow$\{7, 8, 9\} and 0$\leftarrow$2$\leftarrow$5$\leftarrow$\{12, 13, 14\} from Fig.~\ref{fig:framework-overview}.

\header{Results}
The results of Experiment~2 (aimed at answering RQ2) can be found in Table~\ref{tab:replicability}. 
On the \cub{} dataset, CLIP consistently outperforms X-VLM. The biggest relative increase is 124\% for image-to-text in terms of R@10, while the smallest relative increase is 1\% for text-to-image in terms of R@1. Overall, on the text-to-image retrieval task, CLIP outperforms X-VLM by 38\%, and on the image-to-text retrieval task, the relative gain is 70\%.

On \fashion{}, CLIP outperforms X-VLM, too. The smallest relative increase is 9\% for text-to-image in terms of R@1, and the biggest relative increase is 260\% for image-to-text in terms of R@10. In general, on the text-to-image retrieval task, CLIP outperforms X-VLM by 52\%; on the image-to-text retrieval task, the relative gain is 83\%.

Finally, on the \abo{} dataset, CLIP outperforms X-VLM again. The smallest relative increase is 101\% for text-to-image in terms of R@1, and the biggest relative increase is 241\% for image-to-text again in terms of R@10. In general, on the text-to-image retrieval task, CLIP outperforms X-VLM by 139\%; on the image-to-text retrieval task, the relative gain is 190\%.
All in all, CLIP outperforms X-VLM on all three scene-centric datasets. The overall relative gain on \cub{} dataset is 55\%, on \fashion{} dataset -- 101\%. The biggest relative gain of 166\% is obtained on the \abo{} dataset.

\header{Answer to RQ2}
The outcome of Experiment~2 is clear. The original relative performance results obtained on the MS COCO and Flickr30k (Table~\ref{tab:repro}) are only partially replicable to the CUB-200, Fashion200k, and ABO datasets. On the latter datasets CLIP consistently outperforms X-VLM by a large margin, whereas the original scores obtained on the former datasets indicate that X-VLM mostly outperforms CLIP.

\header{Upshot} 
We hypothesize that the failure to replicate the relative results originally reported for scene-centric datasets (viz.\ X-VLM outperforms CLIP) is due to CLIP being pre-trained on more and more diverse image data.
We, therefore, recommend that future work aimed at developing large-scale \ac{CMR} models quantifies and reports the diversity of the training data used.

\begin{table}[t]
\centering
\caption{Results of Experiment 2 (replicability study), using the CUB-200, Fashion200k, and ABO datasets.}
\label{tab:replicability}
\begin{tabular}{l ccc c ccc c ccc}
\toprule
 & \multicolumn{3}{c}{\textbf{Text-to-image}} && \multicolumn{3}{c}{\textbf{Image-to-text}} && \multicolumn{3}{c}{\textbf{Rsum}} \\
\cmidrule{2-4}
\cmidrule{6-8}
\cmidrule{10-12}
\textbf{Model} & 
\multicolumn{1}{c}{\textbf{R@1}} & \multicolumn{1}{c}{\textbf{R@5}} & \multicolumn{1}{c}{\textbf{R@10}} && 
\multicolumn{1}{c}{\textbf{R@1}} & \multicolumn{1}{c}{\textbf{R@5}} & \multicolumn{1}{c}{\textbf{R@10}} && 
\multicolumn{1}{c}{\textbf{t2i}} & \multicolumn{1}{c}{\textbf{i2t}} & \multicolumn{1}{c}{\textbf{total}} \\
\midrule
 & \multicolumn{11}{c}{\textbf{CUB-200}} \\
\midrule
CLIP & 
\textbf{\phantom{0}0.71} & \textbf{\phantom{0}2.38} & \textbf{\phantom{0}4.42} && 
\textbf{\phantom{0}1.23} & \textbf{\phantom{0}3.40} & \textbf{\phantom{0}5.48} && 
\textbf{\phantom{0}7.51} & \textbf{10.11} & \textbf{17.62} \\
X-VLM & 
\phantom{0}0.70 & \phantom{0}2.28 & \phantom{0}2.45 && 
\phantom{0}1.16 & \phantom{0}2.35 & \phantom{0}2.45 && 
\phantom{0}5.43 & \phantom{0}5.96 & 11.39 \\
\midrule
 & \multicolumn{11}{c}{\textbf{Fashion200k}} \\
 \midrule
CLIP & 
\textbf{\phantom{0}3.05} & \textbf{\phantom{0}8.56} & \textbf{12.85} && 
\textbf{\phantom{0}3.43} & \textbf{\phantom{0}9.82} & \textbf{14.56} && 
\textbf{24.46} & \textbf{27.81} & \textbf{52.27} \\
X-VLM & 
\phantom{0}2.80 & \phantom{0}6.62 & \phantom{0}6.70 && 
\phantom{0}1.84 & \phantom{0}3.96 & \phantom{0}4.04 && 
16.12 & 09.84 & 25.96 \\
\midrule
 & \multicolumn{11}{c}{\textbf{ABO}} \\
 \midrule
CLIP & 
\textbf{\phantom{0}6.25} & \textbf{13.90} & \textbf{18.50} && 
\textbf{\phantom{0}7.99} & \textbf{18.96} & \textbf{25.57} && 
\textbf{38.65} & \textbf{52.52} &\textbf{91.17} \\
X-VLM & 
\phantom{0}3.10 & \phantom{0}6.48 & \phantom{0}6.56 && 
\phantom{0}3.20 & \phantom{0}7.42 & \phantom{0}7.50 && 
16.14 & 18.12 & 34.26 \\
\bottomrule
\end{tabular}
\end{table}

\subsection{RQ3: Generalizability} 
To answer RQ3, we compare the relative performance of the selected models on object-centric and scene-centric data. Thus, we compare the relative performance of CLIP and X-VLM on \cub{}, \fashion{}, \abo{} with their relative performance on \coco{} and \flickr{}. We focus on the complete tree from Fig.~\ref{fig:framework-overview}.

\header{Results}
The results of our experiments on the scene-centric datasets are in Table~\ref{tab:repro}; the results that we obtained on the object-centric datasets are in Table~\ref{tab:replicability}. 
On object-centric datasets, CLIP consistently outperforms X-VLM.
However, the situation with scene-centric results is partially the opposite. There, X-VLM outperforms CLIP on the MS COCO dataset.

\header{Answer to RQ3}
Hence, we answer RQ3 by stating that the relative performance results for CLIP and X-VLM that we obtained in our experiments only partially generalize from scene-centric to object-centric datasets. The MS COCO dataset is the odd one out.\footnote{On the GitHub repository for CLIP, several issues have been posted related to the performance of CLIP on the MS COCO dataset. See, e.g., \url{https://github.com/openai/CLIP/issues/115}.}

\header{Upshot}
Given the observed differences in relative performance results for CLIP and X-VLM on scene-centric vs.\ object-centric datasets, we recommend that \ac{CMR} be trained in both scene-centric and object-centric datasets to help improve the generalizability of experimental outcomes.


\section{Discussion \& Conclusions}
\label{sec:conclusions}

We have examined two \ac{SOTA} image-text \ac{CMR} methods, CLIP and X-VLM, by contrasting their performance on two scene-centric datasets (MS COCO and Flicrk30k) and three object-centric datasets (CUB-200, Fashion200k, ABO) in a zero-shot setting.

We focused on the \emph{reproducibility} of the \ac{CMR} results reported in the original publications when evaluated on the selected scene-centric datasets. 
The reported scores were not reproducible for X-VLM when evaluated on the MS COCO and the Flickr30k datasets. For CLIP, we were able to reproduce the scores on the Flickr30k dataset when evaluated using R@5 and R@10.
Conversely, the relative results were reproducible on the MS COCO dataset, for all metrics and tasks, and partially reproducible on the Flickr30k dataset only for the image-to-text task when evaluated on R@1 and R@5. 
We also examined the \emph{replicability} of the \ac{CMR} results using three object-centric datasets. We discovered that the relative results are replicable when we compare the relative performance on the object-centric datasets with the relative scores on the Flickr30k dataset. However, for the MS COCO dataset, the relative outcomes were not replicable. 
And, finally, we explored the generalizability of the obtained results by comparing the models' performance on scene-centric vs.\ object-centric datasets. We observed that the absolute scores obtained when evaluating models on object-centric datasets are much lower than the scores obtained on scene-centric datasets.

Our findings demonstrate that the reproducibility of \ac{CMR} methods on scene-centric datasets is an open problem. Besides, we show that while the majority of \ac{CMR} methods are evaluated on the MS COCO and the Flickr30k datasets, the object-centric datasets represent a challenging and relatively unexplored set of benchmarks.

A limitation of our work is the relatively small number of scene-centric and object-centric datasets used for the evaluation of the models.
Another limitation is that we only considered \ac{CMR} in a zero-shot setting, ignoring, e.g., few-shot scenarios; this limitation did, however, come with the important advantage of reducing the number of experimental design decisions to be made for contrastive experiments.

A promising direction for future work is to include further datasets when contrasting the performance of \ac{CMR} models, both scene-centric and object-centric. In particular, it would be interesting to investigate the models' performance on datasets, e.g., Conceptual Captions~\cite{sharma2018conceptual}, the Flower~\cite{Nilsback08}, and the Cars~\cite{Krause} datasets. 
A natural step after that would be to consider few-shot scenarios.

\subsubsection*{Acknowledgements.}
We thank Paul Groth, Andrew Yates, Thong Nguyen, and Maurits Bleeker for helpful discussions and feedback.

This research was supported by Ahold Delhaize,  and the Hybrid Intelligence Center, a 10-year program funded by the Dutch Ministry of Education, Culture and Science through the Netherlands Organisation for Scientific Research, \url{https://hybrid-intelligence-centre.nl}.

All content represents the opinion of the authors, which is not necessarily shared or endorsed by their respective employers and/or sponsors.

\bibliographystyle{spbasic}
\bibliography{bibliography}

\end{document}